\pdfoutput=1

\documentclass[11pt,a4paper]{article}
\usepackage{bm}
\usepackage{jheppub}
\usepackage{mathrsfs}

\let\de=\partial
\DeclareMathOperator{\tr}{Tr}
\DeclareMathOperator{\dn}{dn}
\DeclareMathOperator{\sn}{sn}
\newcommand{\imag}{\text{i}}
\newcommand{\La}{\mathscr{L}}
\newcommand\dd{\mathop{\text d}\nolimits}
\newcommand{\he}[1]{#1^\dagger}
\newcommand{\vek}[1]{\bm{#1}}
\newcommand{\gr}[1]{\text{#1}}
\newcommand{\skal}[2]{\vek{#1}\cdot\vek{#2}}
\newcommand{\vekt}[2]{\vek{#1}\times\vek{#2}}
\newcommand{\ave}[1]{\langle{#1}\rangle}

\title{Anomalous electrodynamics of neutral pion matter\\
in strong magnetic fields}

\author[a]{Tom\'a\v{s} Brauner}
\author[b]{and Saurabh V.~Kadam}
\affiliation[a]{Faculty of Science and Technology, University of Stavanger,\\
N-4036 Stavanger, Norway}
\affiliation[b]{Indian Institute of Science Education and Research (IISER),\\
Pune 411008, India}
\emailAdd{tomas.brauner@uis.no}
\emailAdd{saurabh.kadam@students.iiserpune.ac.in}

\abstract{The ground state of quantum chromodynamics in sufficiently strong external magnetic fields and at moderate baryon chemical potential is a chiral soliton lattice (CSL) of neutral pions~\cite{Brauner:2016pko}. We investigate the interplay between the CSL structure and dynamical electromagnetic fields. Our main result is that in presence of the CSL background, the two physical photon polarizations and the neutral pion mix, giving rise to two gapped excitations and one gapless mode with a nonrelativistic dispersion relation. The nature of this mode depends on the direction of its propagation, interpolating between a circularly polarized electromagnetic wave~\cite{Yamamoto:2015maz} and a neutral pion surface wave, which in turn arises from the spontaneously broken translation invariance. Quite remarkably, there is a neutral-pion-like mode that remains gapped even in the chiral limit, in seeming contradiction to the Goldstone theorem. Finally, we have a first look at the effect of thermal fluctuations of the CSL, showing that even the soft nonrelativistic excitation does not lead to the Landau-Peierls instability. However, it leads to an anomalous contribution to pressure that scales with temperature and magnetic field as $T^{5/2}(B/f_\pi)^{3/2}$.}

\keywords{Phase Diagram of QCD, Effective Field Theories, Topological States of Matter}

\arxivnumber{}

\begin{document}
 
\maketitle


\section{Introduction}
\label{sec:introduction}

The effects of magnetic field on the phase structure of quantum chromodynamics (QCD) have prompted intensive research activity recently. This wave of interest is motivated by applications to both the relativistic heavy-ion collision experiments and the astrophysics of neutron stars. Many of the theoretical efforts have focused on the properties of matter at zero net baryon number density due to the availability of results from lattice simulations in this regime; see ref.~\cite{Andersen:2014xxa} for a recent review.

The progress in understanding the properties of matter at \emph{nonzero} baryon density has been hindered by the sign problem on the one hand, and by model ambiguities on the other hand. However, an intriguing observation was made a decade ago in ref.~\cite{Son:2007ny}: a combination of an external magnetic field with a nonzero baryon number chemical potential can lead to a spatially modulated condensate of neutral pions as a consequence of the axial anomaly. Such a condensate carries baryon number and can under certain circumstances replace ``ordinary'' (fermionic) nuclear matter as the ground state of QCD.

The exact solution of the equation of motion for the neutral pion condensate has only been found  recently~\cite{Brauner:2016pko}. It turned out that similarly to e.g.~chiral magnets~\cite{Dzyaloshinsky:1964dz} and cholesteric liquid crystals~\cite{Gennes}, the condensate takes the form of a chiral soliton lattice (CSL) when the magnetic field exceeds certain critical value. It was confirmed using a model-independent effective field theory that the CSL is the ground state of QCD within a range of magnetic fields and baryon densities that might occur in the cores of magnetars. Furthermore, the CSL is characterized by a gapless phonon-like excitation that may be interpreted as a Nambu-Goldstone (NG) boson of the spontaneously broken translation invariance.

Since neutral pions interact anomalously with the electromagnetic field, a natural question arises as to what effect the CSL structure has on the propagation of photons. This issue was addressed in ref.~\cite{Yamamoto:2015maz} for the simplest case of the chiral limit where the CSL carries a uniform baryon number density. An intriguing birefringence effect was found. Namely, the dispersion relation of photons propagating along the magnetic field lines depends on their circular polarization: one polarization receives a gap and thus becomes attenuated at low frequencies, whereas the other remains gapless and moreover is softened, acquiring a nonrelativistic, quadratic dispersion relation. The analysis was subsequently further developed in ref.~\cite{Qiu:2016hzd} where more general, possibly time-dependent backgrounds were considered, and the propagation of photons in directions not parallel to the magnetic field was investigated. In another recent paper~\cite{Ozaki:2016vwu}, a simple model for the nonuniform background lattice was analyzed. In such a case, the photon spectrum takes the form of a set of energy bands just like the electron spectrum in solids. Otherwise, its main distinguishing features remain unaltered: one of the photon polarizations is gapped, whereas the other is gapless with a quadratic dispersion relation at low momentum.

All the above-mentioned works are based on the same approximation wherein the chiral background anomalously interacting with the electromagnetic field is treated as fixed. This is justified in condensed-matter analogs such as the topological insulators, which are the actual subject of interest of ref.~\cite{Ozaki:2016vwu}. However, it is not appropriate for the interacting system of neutral pions where the CSL structure emerges spontaneously. As a consequence, the spectrum of the system is expected to contain another soft mode apart from the photons, namely the NG mode of the spontaneously broken translations. It is well known that for a correct low-energy description of a given system, all low-energy degrees of freedom have to be taken into account appropriately.

The goal of the present paper is to fill the gap in the existing analyses and investigate the behavior of the \emph{coupled} system of neutral pion and photon fluctuations of the CSL background. This allows us not only to refine the existing conclusions about the effect of the CSL on the propagation of photons, but also to quantify the effect that dynamical electromagnetic fields have on the CSL and its fluctuations. The results are remarkable. In the absence of coupling between pions and photons, the system will have three gapless excitations: two photon polarizations and the phonon of the CSL. Turning the coupling on gaps two of these excitations and leaves only \emph{one} gapless mode, in spite of spontaneous breaking of spatial translations and the absence of the Higgs mechanism for photons. The only gapless mode is in general a mixture of the two photon polarizations and the neutral pion, which can be disentangled only in special kinematic limits. For instance, for modes propagating along the external magnetic field, unbroken rotational invariance allows one to define helicity and thus clearly distinguish photons from pions. In this limit, one can assert that the CSL phonon receives a gap as a result of its coupling to photons, despite the Goldstone theorem.

The paper is organized as follows. In section~\ref{sec:EFT} we review our basic setup, using the leading-order chiral perturbation theory~\cite{Gasser:1983yg} augmented with the anomalous Wess-Zumino-Witten term. In section~\ref{sec:CSL} we then summarize the most important results of ref.~\cite{Brauner:2016pko} regarding the CSL solution for the neutral pion condensate. Most of the novel material of this paper is contained in section~\ref{sec:spectrum}. We first derive the coupled set of equations of motion for the neutral pion and photon excitations. Closed analytic solutions can be given in several cases, namely in the chiral limit, and for pion-like excitations above the general CSL background, propagating along the external magnetic field. In some other cases we provide approximate solutions in various limits. In section~\ref{subsec:NG} we discuss how the presence of a single gapless excitation is compatible with the Goldstone theorem.

Finally, in section~\ref{sec:finiteT} we initiate the analysis of the thermal excitations in the system. We first show that even the soft, nonrelativistic dispersion of the gapless mode does not lead to the Landau-Peierls instability that would destroy the order parameter at any nonzero temperature. Subsequently, we calculate the leading contribution to pressure at low temperatures. Due to the very same nonrelativistic dispersion, this turns out to scale with temperature as $T^{5/2}(B/f_\pi)^{3/2}$, in contrast to the naively expected $T^4$ scaling for both the photons and the gapless pions (CSL phonons). This previously unnoticed contribution dominates the low-temperature pressure of a pion gas in external magnetic fields.


\section{Low-energy effective theory}
\label{sec:EFT}

The low-energy dynamics of pions is governed by the leading-order chiral perturbation theory ($\chi$PT) Lagrangian
\begin{equation}
\La_\text{$\chi$PT}=\frac{f_\pi^2}4\left[\tr(D_\mu\he\Sigma D^\mu\Sigma)+m_\pi^2\tr(\Sigma+\he\Sigma)\right],
\label{ChPT}
\end{equation}
where $f_\pi$ is the pion decay constant, $m_\pi$ the vacuum pion mass, and $\Sigma$ an $\gr{SU(2)}$-valued field that represents the three pion degrees of freedom. The coupling to the electromagnetic field $A_\mu$ is introduced through a covariant derivative, $D_\mu\Sigma\equiv\de_\mu\Sigma-\imag[Q_\mu,\Sigma]$, where $Q_\mu\equiv A_\mu\frac{\tau_3}2$, and $\tau_3$ is the third Pauli matrix. The effects of the chiral anomaly enter the $\chi$PT Lagrangian at the next-to-leading order of the derivative expansion, and are captured by the Wess-Zumino-Witten term. For two quark flavors, this takes the form~\cite{Son:2007ny}
\begin{equation}
\La_\text{WZW}=j^\mu_\text{B}\left(\frac12A_\mu-A_\mu^\text{B}\right),
\label{WZW}
\end{equation}
where $A^\text{B}_\mu$ is an external gauge field that couples to the baryon number current and $j^\mu_\text{B}$ is the Goldstone-Wilczek current,
\begin{equation}
j^\mu_\text{B}=-\frac1{24\pi^2}\epsilon^{\mu\nu\alpha\beta}\tr\biggl[(\Sigma D_\nu\he\Sigma)(\Sigma D_\alpha\he\Sigma)(\Sigma D_\beta\he\Sigma)+\frac{3\imag}4F_{\nu\alpha}\tau_3(\Sigma D_\beta\he\Sigma+D_\beta\he\Sigma\Sigma)\biggr].
\end{equation}
Finally, the dynamics of the electromagnetic field is described by quantum electrodynamics,
\begin{equation}
\La_\text{QED}=-\frac14F_{\mu\nu}F^{\mu\nu}+\frac1{2\xi}(\de_\mu A^\mu)^2-j^\mu_\text{back}A_\mu,
\label{QED}
\end{equation}
where we defined as usual $F_{\mu\nu}\equiv\de_\mu A_\nu-\de_\nu A_\mu$, added a gauge-fixing term and a coupling to a classical charged background (to be discussed below), represented by the current $j^\mu_\text{back}$.

The total Lagrangian for our system is given by the sum $\La=\La_\text{$\chi$PT}+\La_\text{WZW}+\La_\text{QED}$. Recall that the minimal coupling to an external magnetic field $\vek B_\text{ex}$ gives the \emph{charged} pions a gap proportional to $\sqrt{B_\text{ex}}$~\cite{Andersen:2014xxa}. While it was shown in ref.~\cite{Brauner:2016pko} that the presence of a CSL may reverse this behavior and in sufficiently strong magnetic fields even lead to Bose-Einstein condensation of charged pions, we will assume throughout this paper that the external magnetic field takes such values that the charged pions are sufficiently heavy and therefore can be dropped from the low-energy effective theory. In other words, we shall analyze the dynamics of the coupled system of neutral pions and the electromagnetic field. The master Lagrangian of our system then follows from eqs.~\eqref{ChPT}, \eqref{WZW} and \eqref{QED} by inserting $\Sigma=\exp(\frac\imag{f_\pi}\tau_3\pi^0)$ and setting $A^\text{B}_\mu=(\mu,\vek0)$, where $\mu$ is the baryon chemical potential,
\begin{equation}
\begin{split}
\La={}&\frac12(\de_\mu\pi^0)^2+m_\pi^2f_\pi^2\cos\biggl(\frac{\pi^0}{f_\pi}\biggr)+\frac{\mu C}2\epsilon^{0\mu\nu\alpha}F_{\mu\nu}\de_\alpha\pi^0-\frac C8\pi^0\epsilon^{\mu\nu\alpha\beta}F_{\mu\nu}F_{\alpha\beta}\\
&-\frac14F_{\mu\nu}F^{\mu\nu}+\frac1{2\xi}(\de_\mu A^\mu)^2-j^\mu_\text{back}A_\mu,
\end{split}
\label{master}
\end{equation}
where we used the notation of ref.~\cite{Yamamoto:2015maz} and introduced the abbreviation
\begin{equation}
C\equiv\frac1{4\pi^2f_\pi}.
\end{equation}
The terms in eq.~\eqref{master} containing the Levi-Civita tensor are the contributions of the chiral anomaly. The first one, proportional to $\mu$, is merely a surface term, as is readily seen using the Bianchi identity for the field-strength tensor $F_{\mu\nu}$. It does not affect the field equations of motion, but plays a crucial role in the determination of the ground state~\cite{Brauner:2016pko,Yamamoto:2015maz}. The second term, on the other hand, corresponds to the well-known anomalous coupling of the neutral pion, responsible among others for its two-photon decay in the vacuum. For the record, we also note the nonrelativistic form of the Lagrangian,
\begin{equation}
\begin{split}
\La={}&\frac12\bigl[(\dot\pi^0)^2-(\vek\nabla\pi^0)^2\bigr]+m_\pi^2f_\pi^2\cos\biggl(\frac{\pi^0}{f_\pi}\biggr)+\mu C\skal B\nabla\pi^0+C\pi^0\skal EB\\
&+\frac12(\vek E^2-\vek B^2)+\frac1{2\xi}(\dot\varphi+\skal\nabla A)^2-\rho_\text{back}\varphi+\vek j_\text{back}\cdot\vek A,
\end{split}
\label{masterNR}
\end{equation}
where dots above fields represent time derivatives and we introduced the standard notation $\varphi$, $\vek E$, $\vek B$ for the scalar potential and the electric and magnetic field intensity, respectively. The background charge density and current are denoted naturally as $\rho_\text{back}$ and $\vek j_\text{back}$.

In the following, it will be important to have at hand the equations of motion following from the master Lagrangian~\eqref{master}. These are straightforward to derive, and we therefore just write down their final form in the nonrelativistic notation, assuming the Lorenz gauge $\de_\mu A^\mu=0$,
\begin{equation}
\begin{split}
\ddot\pi^0-\vek\nabla^2\pi^0+m_\pi^2f_\pi\sin\biggl(\frac{\pi^0}{f_\pi}\biggr)&=C\skal EB,\\
\skal\nabla E&=\rho_\text{back}-C\skal B\nabla\pi^0,\\
\vekt\nabla B&=\vek j_\text{back}+\dot{\vek E}+C\vek B\dot\pi^0-C\vekt E\nabla\pi^0.
\end{split}
\label{EoM}
\end{equation}
The first of these equations is the Klein-Gordon equation for the pion field with the Pontryagin density for the electromagnetic field as a source. The second and third then are the usual Maxwell equations with additional contributions coming from the chiral anomaly; they are the basic equations of axion electrodynamics~\cite{Yamamoto:2015maz,Wilczek:1987mv}.

In this paper, we study the electrodynamics of neutral pions in a uniform external magnetic field $\vek B_\text{ex}$. In other words, we are looking for a ground state such that $\vek E=\vek0$ and $\vek B=\vek B_\text{ex}$. Zero electric field requires that the system has to be locally electrically neutral, which is also consistent with the fact that overall electric neutrality is necessary to ensure extensivity of free energy and thus the existence of a thermodynamic limit~\cite{Alford:2002kj}. We therefore assume that in the ground state, the electric charge due to the nonuniform pion field configuration is compensated by the background,
\begin{equation}
\rho_\text{back}=C\vek B_\text{ex}\cdot\ave{\vek\nabla\pi^0},\qquad
\vek j_\text{back}=\vek 0,
\label{background_charge}
\end{equation}
where the angular brackets denote ground state expectation value. Throughout the paper, this background charge density will be assumed to be fixed and unaffected by pion and electromagnetic field fluctuations. This assumption is certainly justified at sufficiently low energies, as we explain in detail in section~\ref{sec:discussion} and appendix~\ref{app:plasma}.


\section{Chiral soliton lattice: an overview}
\label{sec:CSL}

At fixed external magnetic field, the Hamiltonian density of the neutral pion field, following from eqs.~\eqref{ChPT} and~\eqref{WZW}, reads
\begin{equation}
\mathscr{H}=\frac12(\vek\nabla\pi^0)^2-m_\pi^2f_\pi^2\cos\biggl(\frac{\pi^0}{f_\pi}\biggr)-\mu C\skal{B_\text{ex}}\nabla\pi^0.
\label{Hamiltonian}
\end{equation}
It is evident that the anomalous coupling to the magnetic field favors nonuniform field configurations, modulated in the direction of $\vek B_\text{ex}$, chosen here without loss of generality to point along the positive $z$-axis. The one-dimensional equation of motion for such a nonuniform condensate is equivalent to the equation describing a simple pendulum, and can be solved in a closed form in terms of the Jacobi elliptic functions~\cite{Brauner:2016pko,Kishine20151},
\begin{equation}
\cos\frac{\phi(\bar z)}2=\sn(\bar z,k), 
\label{CSL_solution}
\end{equation}
where we introduced the shorthand notation $\phi\equiv\ave{\pi^0}/f_\pi$ and the dimensionless coordinate $\bar z\equiv zm_\pi/k$. Here $k$ is the elliptic modulus, defining the Jacobi elliptic sine.

The above solution represents a lattice of topological solitons, the CSL, with lattice spacing given by
\begin{equation}
\ell=\frac{2kK(k)}{m_{\pi}},
\label{latticespacing}
\end{equation}
where $K(k)$ is the complete elliptic integral of the first kind. The actual value of the elliptic modulus $k$ is determined by minimization of the Hamiltonian, following from eq.~\eqref{Hamiltonian}, and is found to be given implicitly by the condition
\begin{equation}
\frac{E(k)}k=\frac{\mu B_\text{ex}}{16\pi m_\pi f_\pi^2},
\label{condition}
\end{equation}
where $E(k)$ is the complete elliptic integral of the second kind. Since the function $E(k)/k$ is bounded from below by $1$ in the physically acceptable range of $k$ ($0\leq k\leq1$), it follows that the CSL solution exists only above a critical magnetic field $B_\text{CSL}$, equal to
\begin{equation}
B_\text{CSL}=\frac{16\pi m_\pi f_\pi^2}\mu.
\label{BCSL}
\end{equation}
Note that the critical case $B=B_\text{CSL}$ corresponds to the limit $k\to1$ where the Jacobi elliptic functions approach the hyperbolic functions. The solution~\eqref{CSL_solution} then describes an isolated domain wall~\cite{Son:2007ny},
\begin{equation}
\phi_\text{wall}(z)=4\tan^{-1}e^{m_\pi z}.
\label{wall}
\end{equation}
It can be shown that above the critical field $B_\text{CSL}$, the CSL solution is always energetically preferred over the usual QCD vacuum. Moreover, for $\mu\approx m_\text{N}$ where $m_\text{N}$ is the nucleon mass, it is also preferred over nuclear matter~\cite{Brauner:2016pko}.

For illustration, let us insert the physical values $f_\pi\approx92\text{ MeV}$, $m_\pi\approx140\text{ MeV}$, and estimate $\mu=900\text{ MeV}$, that is, close to the expected onset of nuclear matter. We then get from eq.~\eqref{BCSL} that $B_\text{CSL}\approx0.066\text{ GeV}^2$. In the units used here where the Planck constant, speed of light as well as the elementary electric charge is set to one, the conversion to the more usual unit for magnetic field is defined by $1\text{ GeV}^2\approx1.7\times10^{20}\text{ G}$. The critical magnetic field for CSL formation is therefore approximately $10^{19}\text{ G}$, which might be reached in the cores of magnetars~\cite{Andersen:2014xxa}.

While the above-given expressions summarize the \emph{exact} analytic solution of the equation of motion for the neutral pion condensate, it is convenient to have some analytic results in terms of elementary functions. This can be achieved in the chiral limit ($m_\pi\to0$) where the potential term is absent in the Hamiltonian~\eqref{Hamiltonian}. The energy functional is then obviously minimized by a field of a constant gradient,
\begin{equation}
\phi(z)=\frac{\mu B_\text{ex}z}{4\pi^2f_\pi^2}=\frac C{f_\pi}\mu B_\text{ex}z.
\label{CSL_chiral_limit}
\end{equation}
In this case, the soliton ``lattice'' is a linear profile with a uniform charge density. The corresponding ``lattice spacing'' is determined either by taking the limit of eq.~\eqref{latticespacing}, or by using eq.~\eqref{CSL_chiral_limit} directly to find the distance over which the phase $\phi$ increases by $2\pi$,
\begin{equation}
\lim_{m_\pi\to0}\ell=\frac{8\pi^3f_\pi^2}{\mu B_\text{ex}}.
\end{equation}
Below, we will often use the chiral limit as a benchmark case that allows fully analytic calculations. Since the elliptic modulus $k$ depends according to eq.~\eqref{condition} only on the ratio $B_\text{ex}/m_\pi$, the linear profile~\eqref{CSL_chiral_limit} also describes very well the ground state with nonzero pion mass but in sufficiently strong magnetic fields. In fact, the numerical results of ref.~\cite{Brauner:2016pko} show that the chiral limit regime is effectively reached for magnetic fields of just a few (two to three, depending on the precision required) times $B_\text{CSL}$.

Finally, we note that the CSL solution spontaneously breaks translations in the $z$-direction, and we thus expect a phonon-like gapless excitation to appear in the spectrum. This is indeed the case, and the corresponding dispersion relation was found in ref.~\cite{Brauner:2016pko},
\begin{equation}
\omega^2=p_x^2+p_y^2+(1-k^2)\left[\frac{K(k)}{E(k)}\right]^2p_z^2+\mathcal O(p_z^4).
\label{phonondisp}
\end{equation}


\section{Excitation spectrum}
\label{sec:spectrum}

In this section, we analyze in detail the spectrum of excitations above the CSL ground state. We do so by looking for solutions to the equations of motion~\eqref{EoM}, linearized around the CSL background. To that end, we introduce the field fluctuations $\delta\pi^0$, $\delta\vek E$ and $\delta\vek B$ through
\begin{equation}
\pi^0=\ave{\pi^0}+\delta\pi^0,\qquad
\vek E=\delta\vek E,\qquad
\vek B=\vek B_\text{ex}+\delta\vek B.
\end{equation}
Expanding eq.~\eqref{EoM} to first order in these fluctuations and using the equation of motion for the background leads to the set of equations
\begin{equation}
\begin{split}
\delta\ddot\pi^0-\vek\nabla^2\delta\pi^0+m_\pi^2\delta\pi^0\cos\biggl(\frac{\ave{\pi^0}}{f_\pi}\biggr)&=C\vek B_\text{ex}\cdot\delta\vek E,\\
\vek\nabla\cdot\delta\vek E&=-C\delta\vek B\cdot\ave{\vek\nabla\pi^0}-C\vek B_\text{ex}\cdot\vek\nabla\delta\pi^0,\\
\vek\nabla\times\delta\vek B&=\delta\dot{\vek E}+C\vek B_\text{ex}\delta\dot\pi^0-C\delta\vek E\times\ave{\vek\nabla\pi^0}.
\end{split}
\label{EoMlinearized}
\end{equation}
Next we use the fact that the background only depends on the $z$-coordinate to carry out a Fourier transform in the transverse coordinates $x,y$ (denoted collectively with the symbol $\perp$) as well as in time. We introduce the conjugate frequency and transverse momentum variables $\omega$ and $\vek p_\perp$, and the $z$-dependent complex amplitudes $\Pi(z)$, $\vek e(z)$ and $\vek b(z)$, through
\begin{equation}
\begin{split}
\delta\pi^0(\vek r,t)&=\Pi(z)e^{-\imag\omega t}e^{\imag\skal{p_\perp}{r_\perp}},\\
\delta\vek E(\vek r,t)&=\vek e(z)e^{-\imag\omega t}e^{\imag\skal{p_\perp}{r_\perp}},\\
\delta\vek B(\vek r,t)&=\vek b(z)e^{-\imag\omega t}e^{\imag\skal{p_\perp}{r_\perp}}.
\end{split}
\end{equation}
The Bianchi identity $\vekt\nabla E=-\dot{\vek B}$ can then be solved for the magnetic field amplitude. When inserted in eq.~\eqref{EoMlinearized}, one finds that the equation for $\vek\nabla\cdot\delta\vek E$ (the Gauss law) becomes redundant, whereas the rest of eq.~\eqref{EoMlinearized} gives altogether four equations for the unknown functions $\Pi(z)$ and $\vek e(z)$,
\begin{equation}
\begin{split}
(-\de_z^2-\omega^2+\vek p_\perp^2+m_\pi^2\cos\phi)\Pi-CB_\text{ex}e_z&=0,\\
\imag\omega^2CB_\text{ex}\Pi+\vek p_\perp\cdot\de_z\vek e_\perp+\imag(\omega^2-\vek p_\perp^2)e_z&=0,\\
\imag(\de_z^2+\omega^2-p_y^2)e_x+(\imag p_xp_y+\omega Cf_\pi\de_z\phi)e_y+p_x\de_ze_z&=0,\\
(\imag p_xp_y-\omega Cf_\pi\de_z\phi)e_x+\imag(\de_z^2+\omega^2-p_x^2)e_y+p_y\de_ze_z&=0.
\end{split}
\label{EoMmaster}
\end{equation}

These are our master equations that we will use below to find the spectrum of excitations. Before discussing specific cases for the background $\phi(z)$, it is worth pointing out some general features of these equations though. Namely, for modes propagating along the magnetic field ($\vek p_\perp=\vek0$), the equations partially decouple. The first pair reduces to
\begin{equation}
(-\de_z^2-\omega^2+C^2B_\text{ex}^2+m_\pi^2\cos\phi)\Pi=0,
\label{EoMzdirpi}
\end{equation}
augmented with an explicit solution for the longitudinal electric field, $e_z=-CB_\text{ex}\Pi$. This describes a pion, propagating on the CSL background. Since the pion fluctuation induces a fluctuation of local electric charge density, it is naturally accompanied by a fluctuating electric field, but no magnetic field. The term $C^2B_\text{ex}^2$ in eq.~\eqref{EoMzdirpi} is new compared to the situation where the electromagnetic field is fixed to its background value. Since in that case, the pion excitation is always gapless, both in the chiral limit and away from it~\cite{Brauner:2016pko},\footnote{Recall that the neutral pion fluctuation of the CSL simultaneously plays the role of the phonon of the spontaneously broken translation invariance.} we can conclude, without actually solving the equation of motion, that as a result of the feedback from the electromagnetic field fluctuations, the pion receives a gap equal to
\begin{equation}
\omega=CB_\text{ex}=\frac{B_\text{ex}}{4\pi^2f_\pi}.
\end{equation}
Note that since the dispersion relation must be a continuous function of the three-momen\-tum $\vek p$, this result also holds  for other directions of propagation. However, for directions other than $z$, the mode with the above gap does not necessarily behave as the neutral pion; it is in general a nontrivial mixture of oscillating pion and electromagnetic fields.

The second pair of equations in eq.~\eqref{EoMmaster} can for $\vek p_\perp=\vek0$ likewise be rewritten as
\begin{equation}
(-\de_z^2-\omega^2\pm\omega Cf_\pi\de_z\phi)e_\pm=0,
\label{EoMzdire}
\end{equation}
where $e_\pm\equiv e_x\pm\imag e_y$ are the helicity eigenstates of the electric field. This describes a pure transverse electromagnetic wave propagating along the $z$-direction. As noticed in ref.~\cite{Yamamoto:2015maz}, the two helicity eigenstates propagate with different dispersion relations. Without actually solving the equation of motion~\eqref{EoMzdire}, we can see from it that \emph{one} of these eigenstates will always be gapless. The presence of the term linear in $\omega$ indicates that the dispersion relation will be quadratic at low momentum. However, in order to say more about the dispersion relations, we need to specify the background $\phi(z)$.


\subsection{Chiral limit}
\label{subsec:chiral}

We start with the chiral limit, which allows for the most complete analytic treatment without further approximations. As noted above, taking the pion mass to zero is also a reasonably accurate approximation in strong magnetic fields, hence the results obtained below are representative for a large region in the QCD phase diagram in the $\mu$-$B_\text{ex}$ plane.

In the chiral limit, the background~\eqref{CSL_chiral_limit} has a constant gradient, $f_\pi\de_z\phi=\mu CB_\text{ex}$. As a consequence, we can carry out a Fourier transform in the $z$-coordinate as well, upon which the set of equations~\eqref{EoMmaster} simplifies to
\begin{equation}
\begin{split}
(-\omega^2+\vek p^2)\Pi-CB_\text{ex}e_z&=0,\\
\omega^2CB_\text{ex}\Pi+p_z\vek p_\perp\cdot\vek e_\perp+(\omega^2-\vek p_\perp^2)e_z&=0,\\
(\omega^2-p_y^2-p_z^2)e_x+(p_xp_y-\imag\mu\omega C^2B_\text{ex})e_y+p_xp_ze_z&=0,\\
(p_xp_y+\imag\mu\omega C^2B_\text{ex})e_x+(\omega^2-p_x^2-p_z^2)e_y+p_yp_ze_z&=0.
\end{split}
\label{EoMchiral}
\end{equation}
The dispersion relations of the quasiparticle excitations are found by setting the determinant of the coefficients of this set of equations to zero. For a general direction of $\vek p$, this leads to a cubic equation for the squared frequency with a cumbersome analytic solution. It is, however, easy to find the leading terms of the low-momentum expansion for all three different excitations,
\begin{equation}
\begin{split}
\omega_1&=\frac{|\vek p|}{CB_\text{ex}}\sqrt{\vek p_\perp^2+\frac{p_z^2}{\mu^2C^2}}+\mathcal O(\vek p^4),\\
\omega_2&=\mu C^2B_\text{ex}+\mathcal O(\vek p^2),\\
\omega_3&=CB_\text{ex}+\mathcal O(\vek p^2).
\end{split}
\label{dispersions_chiral}
\end{equation}
In other words, there is \emph{one} gapless mode whose dispersion relation is quadratic at low momentum, and anisotropic. The other two modes are gapped.

We would like to briefly contrast this result to those of refs.~\cite{Yamamoto:2015maz,Qiu:2016hzd}. Therein, it was found that for a fixed linear-profile background, the gapless photon mode has a quadratic dispersion relation in the $z$-direction and a linear dispersion relation in other directions. Coupling the photons to the fluctuations of the background therefore has a dramatic effect, making the dispersion quadratic in all directions.

Full dispersion relations can be found for special directions of momentum. For instance, for propagation in the $z$-direction one gets
\begin{equation}
\begin{split}
\omega_{1,2}&=\frac12\Bigl[\mp\mu C^2B_\text{ex}+\sqrt{4p_z^2+(\mu C^2B_\text{ex})^2}\Bigr],\\
\omega_3&=\sqrt{p_z^2+C^2B_\text{ex}^2}.
\end{split}
\end{equation}
It is easy to check using eqs.~\eqref{EoMzdirpi} and \eqref{EoMzdire} that $\omega_3$ corresponds to the pion and $\omega_{1,2}$ to the two circularly polarized photons. On the other hand, for propagation in the transverse plane we find
\begin{equation}
\begin{split}
\omega_{1,3}&=\frac12\Bigl(\mp CB_\text{ex}+\sqrt{4\vek p_\perp^2+C^2B_\text{ex}^2}\Bigr),\\
\omega_2&=\sqrt{\vek p_\perp^2+(\mu C^2B_\text{ex})^2}.
\end{split}
\end{equation}
In this case, both $\omega_1$ and $\omega_3$ constitute a mixture of a pion and an electromagnetic wave polarized linearly in the $z$-direction. The $\omega_2$ mode, on the other hand, corresponds to an electromagnetic wave, polarized eliptically in the $xy$ plane. It is interesting to note that in the low-momentum limit, the gapless mode $\omega_1$ becomes predominantly pion-like. In this limit, \emph{both} photon polarizations are therefore gapped.


\subsection{Single domain wall}
\label{subsec:domainwall}

Before moving away from the chiral limit to the general CSL case, we would like to briefly discuss the domain wall solution~\eqref{wall}, which corresponds to the extreme CSL at the critical magnetic field $B=B_\text{CSL}$~\cite{Son:2007ny}. Since this solution is only of limited interest, we will focus solely on the special case of pions propagating in the $z$-direction.

Substituting from eq.~\eqref{wall}, we find that
\begin{equation}
\cos\phi=1-\frac2{\cosh^2m_\pi z}.
\end{equation}
Introducing a dimensionless coordinate $\tilde z\equiv m_\pi z$ then allows us to rewrite the equation of motion~\eqref{EoMzdirpi} as an eigenvalue problem, $\mathcal H\Pi=\lambda\Pi$, where
\begin{equation}
\mathcal H=-\de_{\tilde z}^2-\frac2{\cosh^2\tilde z},\qquad
\lambda=\frac1{m_\pi^2}(\omega^2-m_\pi^2-C^2B_\text{ex}^2).
\label{wallpi}
\end{equation}
Recognizing $\mathcal H$ as the well-known P\"oschl-Teller Hamiltonian then tells us the spectrum at once. First, there is a single bound state with $\lambda=-1$. In physical units, this solution has the energy $\omega=CB_\text{ex}$ and represents a pion fluctuation localized on the domain wall. Note that such a mode could be expected on general grounds: the \emph{complete} breaking of translation invariance in the $z$-direction by the domain wall leads to a NG mode propagating along the domain wall surface~\cite{Low:2001bw,Watanabe:2014zza}. The peculiarity of our system is that this would-be pion-like NG mode acquires a gap through the interaction with the electromagnetic field. The actual gapless mode appears in the electromagnetic sector of the system.

Second, there is a continuous spectrum of states at $\lambda\geq0$, physically corresponding to energies $\omega\geq\sqrt{m_\pi^2+C^2B_\text{ex}^2}$. These are delocalized states that describe scattering of pions on the domain wall. As is well known, such scattering states are reflectionless for the P\"oschl-Teller potential.


\subsection{General chiral soliton lattice case}
\label{subsec:CSL}

Having discussed the simpler cases of the uniform linear profile and the domain wall, we shall now deal with the spectrum for the general CSL background, given by eq.~\eqref{CSL_solution}. From now on, we will only consider modes propagating in the $z$-direction since as explained above, this allows a clear separation of the pion and photon degrees of freedom, and thus facilitates the physical interpretation of the results. Solutions with a general direction of propagation have to be obtained numerically.

The case of pions is easy to deal with. Indeed, the pion equation of motion, eq.~\eqref{EoMzdirpi}, is identical to the equation of motion for CSL phonons, investigated in ref.~\cite{Brauner:2016pko}, except for the $C^2B_\text{ex}^2$ term. As a consequence, the pion spectrum can be read off eq.~\eqref{phonondisp} upon the corresponding shift,
\begin{equation}
\omega^2=C^2B_\text{ex}^2+(1-k^2)\left[\frac{K(k)}{E(k)}\right]^2p_z^2+\mathcal O(p_z^4).
\end{equation}
Note that as remarked under eq.~\eqref{EoMzdirpi}, the gap in the pion spectrum is independent of the background profile: we found the same result $CB_\text{ex}$ for the linear profile in the chiral limit, for the domain wall as well as for the general CSL solution.

The propagation of photons is governed by eq.~\eqref{EoMzdire}. Below, we provide its approximate analytic solution in two complementary cases: for very strong magnetic fields where $k\to0$ and for close-to-critical magnetic fields ($k\to1$). Finally, we conjecture a closed expression for the photon dispersion relation valid for arbitrary $k$.


\subsubsection{Photons in very strong magnetic fields}

The equation of motion for the circularly polarized photons is given by eq.~\eqref{EoMzdire}. Substituting for the CSL background from eq.~\eqref{CSL_solution}, using some elementary properties of the Jacobi elliptic functions~\cite{Abramowitz:1972}, and introducing the dimensionless coordinate and frequency $\bar z\equiv zm_\pi/k$ and $\bar\omega\equiv\omega k/m_\pi$, it can be cast as
\begin{equation}
\left[-\de_{\bar z}^2-\bar\omega^2\mp\frac{\bar\omega}{2\pi^2}\dn(\bar z,k)\right]e_\pm=0.
\label{photonEoM}
\end{equation}
While we are not able to solve this equation analytically for arbitrary $k$, we can make some progress by using the series expansion of the Jacobi $\dn$ function valid for small $k$,
\begin{equation}
\dn(\bar z,k)\approx1-\frac{k^2}2\sin^2\bar z.
\end{equation}
This represents a well-defined improvement over the result obtained previously in the chiral limit, which would correspond to keeping just the leading $1$ in the expansion. Despite its simplicity, this expression gives, in fact, considerable numerical accuracy over a range of magnetic fields. Using the condition for $k$ in terms of the magnetic field, eq.~\eqref{condition}, one can easily check that the ``potential'' $\dn(\bar z,k)$ is thus approximated with an accuracy of $5\%$ or better for all $\bar z$ for $B/B_\text{CSL}>2$, and with a $1\%$ accuracy for $B/B_\text{CSL}>3$. 

Using the above approximation and introducing some further notation,
\begin{equation}
a\equiv\bar\omega^2\pm\frac{\bar\omega}{2\pi^2}\left(1-\frac{k^2}4\right),\qquad
q\equiv\mp\frac{\bar\omega k^2}{16\pi^2},
\end{equation}
brings the equation of motion~\eqref{photonEoM} into the form of the so-called Mathieu equation,
\begin{equation}
\bigl[\de_{\bar z}^2+(a-2q\cos2\bar z)\bigr]e_\pm=0.
\label{mathieu}
\end{equation}
Since this is equivalent to the Schr\"odinger equation in a periodic potential, the solution can be sought in the Bloch form $e_\pm(\bar z)=e^{\imag\bar p_z\bar z}P(\bar z)$, where $\bar p_z$ is the dimensionless crystal momentum and $P(\bar z)$ is periodic in $\bar z$. According to eq.~(20.3.15) of ref.~\cite{Abramowitz:1972}, the Mathieu ``energy'' $a$ is related to the crystal momentum via
\begin{equation}
a=\bar p_z^2+\frac{q^2}{2(\bar p_z^2-1)}+\mathcal O(q^4).
\end{equation}
While not a fully analytic solution in a closed form, this allows us to determine at least the leading low-momentum behavior of the most interesting, gapless mode in the spectrum. Solving this equation iteratively for $\bar\omega$ as a function of $\bar p_z$ and going back to the physical units, we obtain
\begin{equation}
\omega=\frac{p_z^2}{m_\pi}\frac{2\pi^2 k}{1-\frac{k^2}4}+\mathcal O(p_z^4),
\label{gaplessdisp_smallk}
\end{equation}
where the elliptic modulus $k$ is determined implicitly by eq.~\eqref{condition}.

Eq.~\eqref{gaplessdisp_smallk} is an \emph{exact} solution of the Mathieu equation~\eqref{mathieu} for the dispersion relation of the gapless photon, propagating along the $z$-direction. As such, it is expected to be accurate for medium-to-small $k$. In the next section, we will complement this result by an approximate calculation valid in the opposite limit of $k$ close to one. For the moment, we just observe as a consistency check that the result~\eqref{gaplessdisp_smallk} is compatible with what we previously found for the chiral limit. Indeed, note that in strong fields, the elliptic modulus $k$, as given by eq.~\eqref{condition}, equals asymptotically
\begin{equation}
\frac k{m_\pi}\approx\frac{8\pi^2f_\pi^2}{\mu B_\text{ex}}.
\label{kmpichiral}
\end{equation}
Inserting this in eq.~\eqref{gaplessdisp_smallk} and keeping just the leading term in the expansion in powers of $k$, we reproduce the dispersion relation given in eq.~\eqref{dispersions_chiral}.


\subsubsection{Photons in close-to-critical magnetic fields}

For $k\to1$, the Jacobi elliptic functions take the form of a lattice of widely separated thin domain walls. We can therefore simplify the solution of eq.~\eqref{EoMzdire} by adopting the thin-wall approximation, in which the domain walls are treated as infinitesimally thin. Let us illustrate this approximation on the example of a single isolated domain wall, discussed in section~\ref{subsec:domainwall}. The analog of eq.~\eqref{wallpi} for the photons then reads
\begin{equation}
\left(-\de_{\tilde z}^2-\tilde\omega^2\mp\frac{\tilde\omega}{2\pi^2}\frac1{\cosh\tilde z}\right)e_\pm=0.
\end{equation}
For wavelengths much longer than the thickness of the domain wall, we can approximate the ``potential'' by $1/\cosh\tilde z\to\pi\delta(\tilde z)$. Adding accordingly the other domain walls of the soliton lattice, with spacing $\ell$ given by eq.~\eqref{latticespacing}, then maps the problem onto that of the Dirac comb, known from elementary quantum mechanics and defined by the Hamiltonian~\cite{Fluegge:book}
\begin{equation}
\mathcal H=-\frac1{2m}\frac{\dd^2}{\dd\!\tilde z^2}+\frac\Omega m\sum_{n\in\mathbb{Z}}\delta(\tilde z-na),
\end{equation}
with $a\equiv m_\pi\ell$ and $\Omega\equiv\mp\tilde\omega/(4\pi)$. Using the known spectrum of this Hamiltonian~\cite{Fluegge:book} then gives an implicit solution for the frequency $\omega$ as a function of momentum $p_z$,
\begin{equation}
\cos(p_z\ell)=\cos(\omega\ell)\mp\frac1{4\pi}\sin(\omega\ell).
\end{equation}
Note that in this thin-wall approximation, the sole dependence on the external magnetic field $\vek B_\text{ex}$ comes through the lattice spacing $\ell$. One of the photon helicities is gapless as expected, with the low-momentum dispersion relation given by
\begin{equation}
\omega=4\pi kK(k)\frac{p_z^2}{m_\pi}+\mathcal O(p_z^4).
\label{photon_gapless}
\end{equation}
The other helicity is gapped with the energy
\begin{equation}
\omega=\frac{m_\pi}{kK(k)}\tan^{-1}\frac1{4\pi}+\mathcal O(p_z^2).
\label{photon_gapped}
\end{equation}

\begin{figure}
\begin{center}
\includegraphics[scale=1.2]{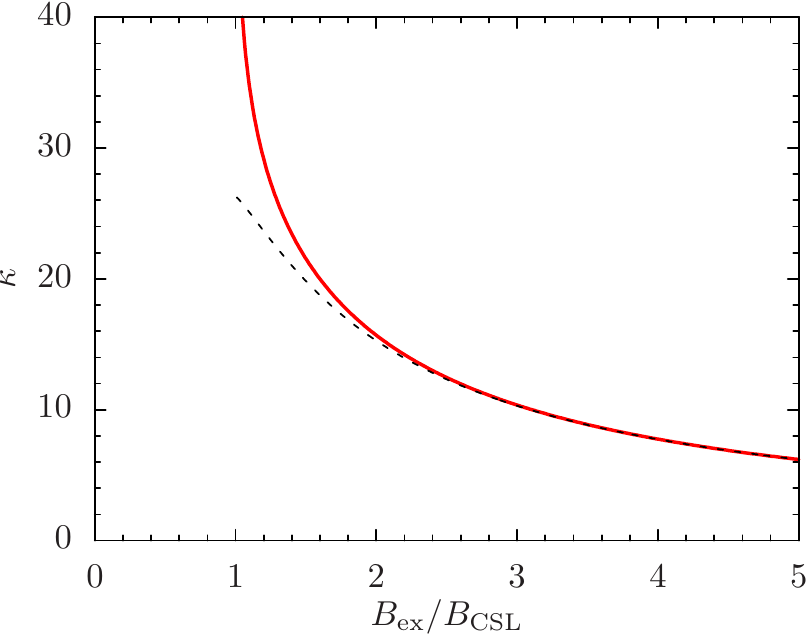}
\end{center}
\caption{The dimensionless factor $\kappa$ in the gapless photon dispersion relation, $\omega=\kappa p_z^2/m_\pi$, as a function of the magnetic field. The elliptic modulus $k$ is determined by the magnetic field implicitly by eq.~\eqref{condition}. The solid red line corresponds to eq.~\eqref{photon_gapless}, conjectured to hold across the whole range of $k$. The dashed black line corresponds to eq.~\eqref{gaplessdisp_smallk}, obtained in the limit $k\to 0$. The divergence of the coefficient $\kappa$ in the limit $B_\text{ex}\to B_\text{CSL}$ indicates that the dispersion relation is no longer quadratic at low momentum in this limit. Indeed, in presence of a single isolated domain wall, there are photons propagating in the bulk with the usual relativistic dispersion relation.}
\label{fig:gaplessdisp}
\end{figure}

The results~\eqref{gaplessdisp_smallk} and~\eqref{photon_gapless} for the gapless photon dispersion relation, obtained in the opposite limits of $k\to0$ and $k\to1$, agree in their power expansions in $k$ up to corrections of order $k^5$. This indicates that there might be an (albeit approximate) expression for the dispersion relation, valid across the whole range $0\leq k\leq1$. To get such an expression, we return to eq.~\eqref{photonEoM}. Provided that the wavelength of the photon is long enough, we can replace the Jacobi elliptic function $\dn(\bar z,k)$ with its average value~\cite{Abramowitz:1972},
\begin{equation}
\dn(\bar z,k)\to\frac\pi{2K(k)},
\end{equation}
upon which eq.~\eqref{photonEoM} can be Fourier-transformed, leading immediately to
\begin{equation}
\bar p_z^2-\bar\omega^2\mp\frac{\bar\omega}{4\pi K(k)}=0.
\end{equation}
Remarkably, this reproduces exactly the dispersion relation~\eqref{photon_gapless}, obtained using a completely different approach. Although the above argument is far from rigorous, it leads us to conjecture that the dispersion relation of the gapless photon is given exactly or accurately by eq.~\eqref{photon_gapless} for an arbitrary value of the magnetic field (and thus $k$). A comparison of different approximations to the gapless photon dispersion is shown in figure~\ref{fig:gaplessdisp}.

Finally, the same approximation gives the energy of the gapped photon at $p_z=0$ as $\omega=m_\pi/[4\pi kK(k)]$. This is not exactly equal to the result obtained using the thin-wall approximation, eq.~\eqref{photon_gapped}. However, the dependence of the two results on $k$ is the same and the numerical values of the prefactor only differ by about $0.2\%$. It it interesting to note that the above expression has the correct asymptotic behavior in strong magnetic fields, or in the chiral limit. Indeed, it is easy to check using eq.~\eqref{kmpichiral} that for $k\to0$ it reproduces the gap of the $\omega_2$ mode in eq.~\eqref{dispersions_chiral}.


\subsection{Identification of the Nambu-Goldstone boson}
\label{subsec:NG}

Our results for the dispersion relations of the various excitations were obtained using model-independent effective field theory, and therefore apply to any physical system described by axion electrodynamics coupled to a light neutral scalar. Since the effective field theory framework is largely based on symmetry, it would be instructive to gain some insight into the results in relation to spontaneous breaking of the symmetries of the system. This is indeed known to have nontrivial consequences in case of spacetime symmetries. First, some of the symmetries may be redundant and thus not give rise to independent NG modes in the spectrum~\cite{Watanabe:2013iia,Hayata:2013vfa}. Second, in presence of topological solitons, a central charge may appear in the Lie algebra of the symmetry generators upon quantization, leading to quadratic dispersion relations and a further reduction of the number of NG modes~\cite{Watanabe:2014pea,Kobayashi:2014xua}.

Let us start by listing the symmetries of the system as defined by the Lagran\-gi\-an~\eqref{master}. Once the charged pions are discarded, all that is left of the chiral symmetry of QCD is an internal $\gr{U(1)}$ symmetry under which $\pi^0$ shifts by a constant. This is an approximate symmetry for a nonzero pion mass $m_\pi$ that becomes exact in the chiral limit. It is complemented by the gauge invariance of quantum electrodynamics. Finally, the Poincar\'e symmetry of relativistic field theory is reduced by the magnetic field $\vek B_\text{ex}$ to spacetime translations, boosts in the $z$-direction and rotations around the $z$-axis.

The CSL ground state breaks spontaneously translations and boosts in the $z$-direction as well as the internal $\gr{U(1)}$ symmetry. The $z$-boost is redundant with the $z$-translation and thus does not give rise to an independent NG mode~\cite{Watanabe:2013iia}. The low-energy spectrum of the system is therefore determined by the spontaneous breaking of the internal $\gr{U(1)}$ symmetry and of translations in the $z$-direction. In the chiral limit, a ``diagonal'' $\gr{U(1)}$ subgroup is left unbroken by the linear profile~\eqref{CSL_chiral_limit}, under which the $\pi^0$ field and the $z$-coordinate shift simultaneously. Off the chiral limit, only a discrete subgroup thereof is left unbroken by the CSL background~\eqref{CSL_solution}. However, only the translations then represent an exact symmetry in the first place. Therefore, there is \emph{one} exact symmetry for any $m_\pi$ that is spontaneously broken and gives rise to a NG mode.

To understand the connection of the anticipated NG mode to the gapless mode found through a direct computation, we calculate the Noether currents of the two symmetries in question. First, an internal $\gr{U(1)}$ symmetry transformation shifts the Lagrangian by a surface term due to the presence of the anomalous $\pi^0\skal EB$ interaction. This affects the form of the corresponding current,
\begin{equation}
j^\mu=\de^\mu\pi^0+\frac C4\epsilon^{\mu\nu\alpha\beta}A_\nu F_{\alpha\beta}.
\end{equation}
This current contains a term linear in $\pi^0$, and we therefore expect the $\pi^0$ field to couple to the one-particle NG state. Note that thanks to the background magnetic field, the current also contains a term linear in $A_\mu$, in particular the corresponding charge density $j^0$ gets a contribution proportional to $A_z$. However, this contribution is gauge-dependent and thus should not be assigned a definite physical meaning.

The translational invariance of the system implies the existence of a conserved energy-momentum tensor. The symmetric energy-momentum tensor that can be obtained for instance by varying the action with respect to a background metric, reads
\begin{equation}
\begin{split}
\Theta^{\mu\nu}={}&\de^\mu\pi^0\de^\nu\pi^0-F^{\mu\alpha}F^\nu_{\phantom\nu\alpha}-g^{\mu\nu}\La\\
={}&\de^\mu\pi^0\de^\nu\pi^0-g^{\mu\nu}\left[\frac12(\de_\alpha\pi^0)^2+m_\pi^2f_\pi^2\cos\left(\frac{\pi^0}{f_\pi}\right)\right]\\
&-F^{\mu\alpha}F^\nu_{\phantom\nu\alpha}+\frac14g^{\mu\nu}F_{\alpha\beta}F^{\alpha\beta}+\frac C8g^{\mu\nu}\pi^0\epsilon^{\alpha\beta\gamma\delta}F_{\alpha\beta}F_{\gamma\delta},
\end{split}
\end{equation}
where we dropped the surface term in the Lagrangian~\eqref{master} that does not contribute to the action for the order parameter fluctuations as well as the gauge-fixing and the background current terms. The generator of translations in the $z$-direction is the corresponding momentum density,
\begin{equation}
P_z=\Theta^{0z}=-\dot\pi^0\de_z\pi^0-F^{0\alpha}F^z_{\phantom z\alpha}.
\end{equation}
This again contains a term linear in the fluctuation $\delta\pi^0$ provided that $\ave{\de_z\pi^0}\neq0$, which is exactly the condition for spontaneous breaking of translations. There is obviously no term linear in the electromagnetic field fluctuation in this case.

Altogether, we conclude that the spectrum of the system should contain one NG boson that couples to the $\pi^0$ field. This is indeed consistent with the explicit calculation carried out above. However, the realization of the Goldstone theorem is still nontrivial. Namely, recall that for modes propagating along the $z$-direction, we found that the pion and electromagnetic fields decouple and the pion mode is gapped. The $\pi^0$ field nevertheless has an overlap with the gapless state in the spectrum for all \emph{other} directions of momentum than along the $z$-axis. This is sufficient to saturate the Goldstone commutator where the whole phase space is integrated over. Finally, it is remarkable that apart from the mode predicted by the Goldstone theorem, there are no other gapless modes in the spectrum in spite of the gauge invariance of the system. The single gapless mode does contain an admixture of the electromagnetic field for all directions of propagation though.


\section{Effects of thermal fluctuations}
\label{sec:finiteT}

Now that we know the spectrum of excitations of the CSL in presence of dynamical electromagnetic fields, we shall in this section briefly discuss the effects of these excitations at nonzero temperature. This is important, among others, for establishing the stability of the CSL. Indeed, it is well known that some ordered states with one-dimensional modulation are unstable with respect to thermal fluctuations in the transverse directions, destroying the long-range order at an arbitrarily low but nonzero temperature~\cite{Lee:2015bva,Hidaka:2015xza}. It was argued in ref.~\cite{Brauner:2016pko} that the CSL is safe from such an instability due to the linear dispersion relation of the phonon excitation~\eqref{phonondisp}, which makes the thermal fluctuations essentially harmless. However, as we saw in the previous section, including the coupling to dynamical electromagnetic fields makes the dispersion relation of the softest mode quadratic in all directions, providing a much larger density of states and in turn a substantial enhancement of infrared fluctuations. The argument therefore has to be carefully revisited.

In this section, we restrict ourselves for simplicity to the chiral limit where a simple analytic expression for the background CSL profile~\eqref{CSL_chiral_limit} is available. We go back to the Lagrangian~\eqref{master} and expand it to second order in the fluctuations of the pion and electromagnetic fields, $\delta\pi^0$ and $\delta A_\mu$. Formally, this bilinear Lagrangian can be cast as
\begin{equation}
\La_\text{bilin}=\frac12
\begin{pmatrix}
\delta\pi^0 & \delta A^\mu
\end{pmatrix}
\mathscr{D}^{-1}_{\mu\nu}
\begin{pmatrix}
\delta\pi^0\\
\delta A^\nu
\end{pmatrix},
\end{equation}
where the $5\times5$ inverse matrix propagator takes, upon a Fourier transform, the form
\begin{equation}
\renewcommand{\arraystretch}{1.3}
\mathscr{D}^{-1}_{\mu\nu}(p)=
\left(\begin{array}{c|c}
p^2 & \frac{\imag C}2\epsilon_{\mu\nu\alpha\beta}\ave{F^{\alpha\beta}}p^\mu\\
\hline
\frac{\imag C}2\epsilon_{\mu\nu\alpha\beta}\ave{F^{\alpha\beta}}p^\nu & -g_{\mu\nu}p^2+\bigl(1+\frac1\xi\bigr)p_\mu p_\nu+\imag C\epsilon_{\mu\nu\alpha\beta}\ave{\de^\alpha\pi^0}p^\beta
\end{array}\right).
\label{matpropagator}
\end{equation}

Whether or not the thermal fluctuations destroy the order parameter can be judged by looking at the two-point correlator of the field that defines this order parameter~\cite{Coleman:1973ci}, in this case $\pi^0$. At nonzero temperature, the infrared fluctuations are dominated by the zero Matsubara mode. Setting frequency to zero and inverting the matrix~\eqref{matpropagator}, we then get for the $\pi^0\pi^0$ correlator in momentum space the expression
\begin{equation}
\mathscr{D}_{\pi^0\pi^0}(\omega=0)=-\frac{(\vek p^2)^2+(\mu C^2B_\text{ex})^2\vek p_\perp^2}{\vek p^2\bigl[(\vek p^2)^2+(\mu C^2B_\text{ex})^2\vek p_\perp^2+C^2B_\text{ex}^2p_z^2\bigr]}.
\end{equation}
This obviously gives an infrared-finite correlation function upon Fourier transforming to the coordinate space. We therefore conclude that the thermal excitations do not give rise to an instability of our system in spite of their quadratic dispersion relation.

In order to assess the effect of the thermal fluctuations on the thermodynamics of the system, we next evaluate their leading, one-loop contribution to the pressure. To that end, we need to evaluate the thermal sum-integral of the logarithm of the determinant of $\mathscr D^{-1}_{\mu\nu}(p)$. First we note that the sole dependence of the determinant on the gauge-fixing parameter $\xi$ comes through an overall prefactor $1/\xi$, and hence the pressure is gauge-independent as it should be. In addition, the determinant carries a trivial factor $(p^2)^2=(\omega^2-\vek p^2)^2$, which is cancelled by the contribution of ghost degrees of freedom. The thermal part of the pressure\footnote{We shall not discuss the zero-temperature one-loop contributions to pressure from the chemical potential and magnetic field here.} is then given by
\begin{equation}
P=-T\sum_{i=1}^3\int\frac{\dd^3\!\vek p}{(2\pi)^3}\log\bigl[1-e^{-\beta\omega_i(\vek p)}\bigr],
\end{equation}
where the functions $\omega_i(\vek p)$ coincide with the dispersion relations of the three physical excitations, derived using the equations of motion in section~\ref{subsec:chiral}. This is of course expected; the argument of this section, based on the evaluation of the fluctuation determinant, therefore serves merely as an independent verification of the results for the dispersion relations, obtained before.

\begin{figure}
\begin{center}
\includegraphics[scale=1.2]{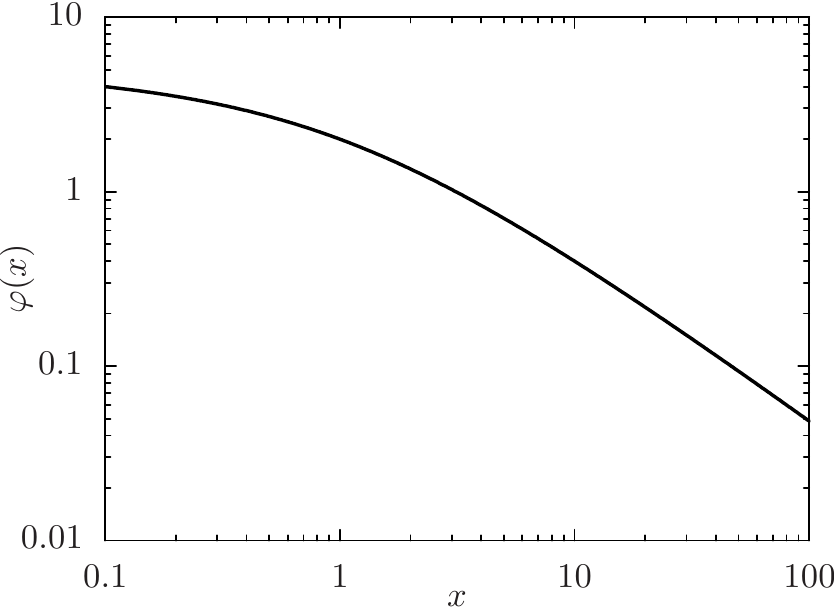}
\end{center}
\caption{Numerical values of the anisotropy factor $\varphi(x)$, defined by eq.~\eqref{anisotropy}. In physical units and for the physical value $f_\pi\approx92\text{ MeV}$, the argument $x=1/(\mu C)$ corresponds roughly to $3600\text{ MeV}/\mu$. Thus, only values considerably larger than one are realistic.}
\label{fig:varphi}
\end{figure}

Due to the complicated form of the dispersion relations for a general direction of propagation, it seems impossible to give a closed expression for the pressure even in the chiral limit. However, noting that two of the excitations are gapped, see eq.~\eqref{dispersions_chiral}, we can obtain a closed \emph{approximate} expression for the pressure, valid at sufficiently low temperatures. As long as $T\ll\min\{\mu C^2B_\text{ex},CB_\text{ex}\}$, only the gapless mode $\omega_1$ contributes significantly. Approximating furthermore its dispersion with the leading $\mathcal O(\vek p^2)$ term leads to an integral that is easily calculable, giving
\begin{equation}
P=\frac{\zeta(\frac52)}{16\pi^{3/2}}(CB_\text{ex})^{3/2}T^{5/2}\varphi(\textstyle\frac1{\mu C}).
\label{pressure}
\end{equation}
The sole dependence of pressure on the chemical potential comes from the anisotropy factor
\begin{equation}
\varphi(x)\equiv\int_0^\pi\frac{\sin\theta\dd\!\theta}{(\sin^2\theta+x^2\cos^2\theta)^{3/4}}.
\label{anisotropy}
\end{equation}
The function $\varphi(x)$ is bounded on the real line, and is plotted numerically in figure~\ref{fig:varphi}. For certain special values of the argument it can be evaluated in terms of elementary functions, thus: $\varphi(0)=2^{3/2}K(2^{-1/2})\approx5.24$, $\varphi(1)=2$, and $\varphi(\infty)=0$.

Note that the expression for the pressure in eq.~\eqref{pressure} vanishes in the limit $\mu\to0$. This indicates merely that our approximation is not appropriate in this limit. First of all, there is no CSL in the ground state at zero chemical potential. It turns out that the anomalous coupling of neutral pions to photons still leads to their mixing at $\mu=0$ as long as an external magnetic field is present. However, the resulting dispersion relations scale differently with momentum and, likewise, the pressure scales differently with temperature and magnetic field than eq.~\eqref{pressure} would suggest.\footnote{Work in progress.}


\section{Summary and discussion}
\label{sec:discussion}

In ref.~\cite{Brauner:2016pko} it was shown that the ground state of QCD at moderate baryon chemical potential and in sufficiently strong magnetic fields contains a neutral pion condensate forming a soliton lattice. In this paper, we investigated the spectrum of excitations above such CSL state, taking into account the coupling of neutral pions to dynamical electromagnetic fields. We found analytic expressions for the dispersion relations at low momentum in case of propagation along the magnetic field,
\begin{equation}
\begin{split}
\omega_1&=4\pi kK(k)\frac{p_z^2}{m_\pi}+\mathcal O(p_z^4),\\
\omega_2&=\frac{m_\pi}{4\pi kK(k)}+\mathcal O(p_z^2),\\
\omega_3&=CB_\text{ex}+\mathcal O(p_z^2),
\end{split}
\end{equation}
where the elliptic modulus $k$ is determined implicitly in terms of the external magnetic field $\vek B_\text{ex}$ by eq.~\eqref{condition}. The modes $\omega_{1,2}$ describe the two circular polarizations of the photon, whereas $\omega_3$ corresponds to the neutral pion. More complete solution including other directions of propagation is possible in the chiral limit, or equivalently in the limit of very strong magnetic fields, and is given in eq.~\eqref{dispersions_chiral}. In general, the three excitation modes are admixtures of the pion and photon degrees of freedom. One of the modes is always gapless with a quadratic dispersion relation whereas the other two modes are gapped.

Our second main result is an explicit expression for the pressure of the system~\eqref{pressure}, valid in the chiral limit and at sufficiently low temperatures such that the contributions of the gapped modes are exponentially suppressed. As a result of the quadratic dispersion relation of the gapless mode, the pressure features an anomalous dependence on temperature, being proportional to $(CB_\text{ex})^{3/2}T^{5/2}$. The chiral anomaly thus changes the thermodynamics of the system qualitatively: in contrast to the naively expected black-body-like scaling $\sim T^4$, the pressure becomes enhanced at low temperatures as a consequence of the high density of states for the gapless mode.  This previously unnoticed contribution dominates the low-temperature pressure of a pion gas in external magnetic fields and at nonzero baryon chemical potential.

\begin{figure}
\begin{center}
\includegraphics[scale=1.2]{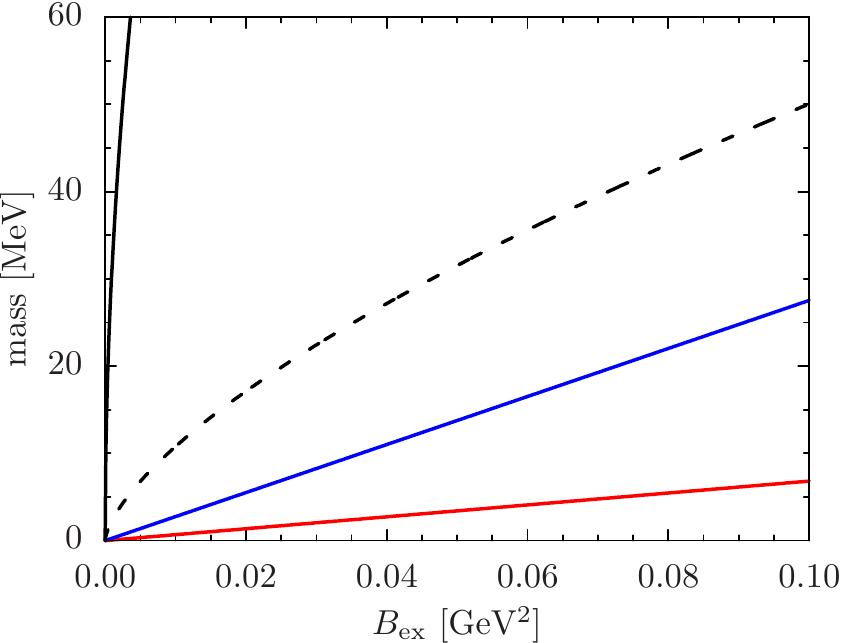}
\end{center}
\caption{Comparison of various energy scales in our system as a function of the external magnetic field at $\mu=900\text{ MeV}$. The red and blue lines correspond to the gapped modes listed in eq.~\eqref{dispersions_chiral}: $\omega_2=\mu C^2B_\text{ex}$ and $\omega_3=CB_\text{ex}$, respectively. The solid black line represents the mass of the charged pion in the chiral limit, $m_{\pi^\pm}=\sqrt{B_\text{ex}}$. The dashed black line stands for the frequency of plasma oscillations of the background, eq.~\eqref{plasma_frequency}. All numerical values were obtained using $f_\pi\approx92\text{ MeV}$.}
\label{fig:masses}
\end{figure}

We would like to append a remark concerning our effective field theory setup. While the basic framework is based solely on symmetry and thus is model-independent, we did make an approximation by neglecting the dynamics of charged pions as well as the oscillations of the charged background, required to render the system electrically neutral as a whole. We would now like to justify this approximation and at the same time illustrate some of our results numerically. For the sake of simplicity, we will restrict ourselves to the chiral limit. There, the low-energy spectrum of the system contains four different energy scales, corresponding to the gaps of various excitation branches. These are the two gapped neutral-pion-photon mixtures, the charged pion, and the plasma oscillation of the background (see appendix~\ref{app:plasma} for a discussion of the latter). Collecting the results displayed in eqs.~\eqref{dispersions_chiral} and~\eqref{plasma_frequency}, their gaps are respectively given by
\begin{equation}
\omega_2=\mu C^2B_\text{ex},\qquad
\omega_3=CB_\text{ex},\qquad
\omega_{\pi^\pm}=\sqrt{B_\text{ex}},\qquad
\omega_\text{pl}=\left[\frac{\mu(CB_\text{ex})^2}{\pi\sqrt3}\right]^{1/3}.
\end{equation}
(The gap of the charged pion follows from the standard Landau level quantization~\cite{Andersen:2014xxa}.) The numerical values of these gaps are shown in figure~\ref{fig:masses} for a realistic range of magnetic fields, using $\mu=900\text{ MeV}$ for illustration.\footnote{While fine details might change, the basic pattern would remain the same for other values of $\mu$ or for a nonzero but sufficiently light pion mass $m_\pi$.} This confirms that our setup is consistent: the two neglected modes are considerably heavier than the two gapped modes in the neutral-pion-photon sector. The range of validity of our effective theory is set by the plasma oscillation frequency~\eqref{plasma_frequency}. The red curve in figure~\ref{fig:masses} at the same time illustrates the range of validity of the low-temperature approximation, leading to the expression for the pressure~\eqref{pressure}.

Finally, we add a comment on the presence of a quadratic dispersion relation of the photon, first observed in ref.~\cite{Yamamoto:2015maz}. Regardless of the concrete dynamics responsible for this effect, it is an interesting question per se to ask under what circumstances the photon \emph{can} acquire a quadratic dispersion. Barring accidental cancellations in the spatial part of the photon's kinetic term, the only possibility seems to be the presence of a term with a single time derivative, schematically $A_i\de_0A_j$, in the bilinear part of the photon Lagrangian. It turns out that under the requirement of full translation invariance, there is only one such term that is consistent with $\gr{U(1)}$ gauge invariance, namely the Chern-Simons term in two spatial dimensions~\cite{Brauner:2014ata}. This term is automatically Lorentz-invariant and gives the photon a mass rather than making it nonrelativistic. Therefore, the \emph{only} way to give the photon a quadratic dispersion relation is to break translations, explicitly or spontaneously. Spontaneous breaking of translations in one direction as in the CSL scenario discussed in this paper provides a particular, minimal mechanism to accomplish this goal.


\acknowledgments
We are indebted to Xu-Guang Huang and Naoki Yamamoto for illuminating discussions and correspondence. S.K.~acknowledges the hospitality of the Department of Mathematics and Natural Sciences, University of Stavanger, where this work was completed.


\appendix

\section{Plasma oscillations of the background}
\label{app:plasma}

In this appendix, we provide some details behind our assumption that the charged background, compensating the electric charge of the CSL itself, does not affect the dynamics of the low-energy excitations discussed in this paper. The actual composition of this background may depend on the concrete (astro)physical environment in which the CSL of neutral pions is realized: it may consist for instance of charged pions, electrons or nucleons~\cite{Son:2007ny}. For the sake of an estimate, we will assume that the background is composed of massless spin-$1/2$ particles of unit electric charge, coupled minimally to the electromagnetic field. Also, we will focus on the chiral limit where the CSL carries a uniform charge density.

In presence of an electromagnetic field, a gas of charged massless particles undergoes plasma oscillations, characterized in the long-wavelength limit by the frequency~\cite{Kapusta:2006kg}
\begin{equation}
\omega_\text{pl}=\sqrt{\frac{\rho_\text{back}}{\mu_\text{back}}},
\end{equation}
where $\mu_\text{back}$ is the chemical potential of the gas and $\rho_\text{back}$ its charge density. (Recall that we use units in which the elementary electric charge is set to one.) Neglecting the background dynamics is therefore justified for energies well below this frequency, which sets the range of validity of our effective theory.

The charge density needed to neutralize the CSL is given by eq.~\eqref{background_charge},
\begin{equation}
\rho_\text{back}=\mu(CB_\text{ex})^2.
\end{equation}
For a gas of massless spin-$1/2$ Dirac fermions, the density is related to the chemical potential via $\rho_\text{back}=\mu_\text{back}^3/(3\pi^2)$. We thus obtain for the plasma frequency
\begin{equation}
\omega_\text{pl}=\frac{\mu_\text{back}}{\pi\sqrt3}=\left(\frac{\rho_\text{back}}{\pi\sqrt3}\right)^{1/3}=\left[\frac{\mu(CB_\text{ex})^2}{\pi\sqrt3}\right]^{1/3}.
\label{plasma_frequency}
\end{equation}
This expression serves as a basis for our discussion of scale separation in section~\ref{sec:discussion}.


\bibliographystyle{JHEP}
\bibliography{references}

\end{document}